\documentclass[11pt]{revtex4}
\usepackage[pdftex]{graphicx}
\usepackage{amsmath}
\usepackage{amsfonts}
\usepackage{amssymb}
\begin{document}
\title{Massless solutions for families of spinors for the toy model~\cite{DNFAM} in $d=(1+5)$} 
\author{M. \v Stimulak, A. Zalo\v znik, N.S. Manko\v c Bor\v stnik \\
Department of Physics, FMF, University of Ljubljana,\\
Jadranska 19, SI-1000 Ljubljana, Slovenia
}

\begin{abstract} 
Massless solutions for four coupled first order differential equations  for functions
representing  families of spinors on the infinite disc curled into  an almost $S^2$ 
are presented and normalizability conditions discussed. Spinors interact with two 
kinds of spin connection fields of particular coordinate dependence~\cite{DNFAM}. 
\end{abstract}
\maketitle
\section{Introduction}
\label{introduction}

We present in this contribution massless solutions of equations of motion for families 
of spinors on an infinite disc curled on an almost $S^{2}$. Spinors interact with two kinds 
of the spin connection fields. The action leads~\cite{DNFAM} to two decoupled groups of four 
coupled first order differential equations, each for four functions. The two groups are not 
really unconnected: There are three spin-connection fields which appear in both groups  
of four equations. The strengths of these and other spin connection fields determine possible 
massless solutions.  

Below four coupled first order differential equations for one of the two groups are 
presented~\cite{DNFAM} 
\begin{eqnarray}
\label{mass12}
&&-if \bigl\{ [(\frac{\partial}{\partial\rho} - \frac{n}{\rho}) 
           - \frac{1}{2f} \frac{\partial f}{\partial\rho} 
           (1-2F_{56}-2\tilde{F}_{56}-2\tilde{F}^{\ominus 3})]\,  \mathcal{A}_n^I \nonumber\\
           &-& \frac{1}{2f} \frac{\partial f}{\partial\rho}
           \, 2\tilde{F}^{\ominus \boxplus}\, \mathcal{A}_n^{II}
    \bigr\} 
    + m \, \mathcal{B}_{n+1}^I = 0\,,\nonumber\\
&&-if \bigl\{ [(\frac{\partial}{\partial\rho} + \frac{n+1}{\rho}) 
           - \frac{1}{2f} \frac{\partial f}{\partial\rho} 
                (1+2F_{56}-2\tilde{F}_{56} - 2\tilde{F}^{\ominus 3})] \, \mathcal{B}_{n+1}^{I} \nonumber\\
           &-& \frac{1}{2f} \frac{\partial f}{\partial\rho} 
           \, 2\tilde{F}^{\ominus \boxplus}\,  \mathcal{B}_{n+1}^{II}
   \bigr\} + m \, \mathcal{A}_{n}^{I} = 0\,,
\end{eqnarray}
\begin{eqnarray}
&&-if \bigl\{ [(\frac{\partial}{\partial\rho} - \frac{n}{\rho}) 
           - \frac{1}{2f} \frac{\partial f}{\partial\rho} 
            (1-2F_{56}-2\tilde{F}_{56}+ 2\tilde{F}^{\ominus 3})] \, \mathcal{A}_n^{II} \nonumber\\
           &-& \frac{1}{2f} \frac{\partial f}{\partial\rho}
           \, 2\tilde{F}^{\ominus \boxminus}\, \mathcal{A}_n^{I}
    \bigr\} + m \,\mathcal{B}_{n+1}^{II} = 0\,,\nonumber\\
&&-if \bigl\{ [(\frac{\partial}{\partial\rho} + \frac{n+1}{\rho}) 
           - \frac{1}{2f} \frac{\partial f}{\partial\rho} 
              (1+2F_{56}-2\tilde{F}_{56}+2\tilde{F}^{\ominus 3})] \, \mathcal{B}_{n+1}^{II} \nonumber\\
           &-& \frac{1}{2f} \frac{\partial f}{\partial\rho} 
              \, 2\tilde{F}^{\ominus \boxminus}\, \mathcal{B}_{n+1}^{I}
    \bigl\} + m \,\mathcal{A}_{n}^{II} = 0\,.\nonumber
\end{eqnarray}
The parameters ($F_{56}\,$, $\tilde{F}_{56}\,$, $\tilde{F}^{\ominus \boxminus}\,$, 
$\tilde{F}^{\ominus \boxminus}\,$, $\tilde{F}^{\ominus 3}\,$) are assumed to be free.

For the massless case these four coupled equations decouple into equations for $\mathcal{A}_n^{I, II}$ 
\begin{eqnarray}
\label{massless12A}
&&[(\frac{\partial}{\partial\rho} - \frac{n}{\rho}) 
           - \frac{1}{2f} \frac{\partial f}{\partial\rho} 
           (1-2F_{56}-2\tilde{F}_{56}- 2\tilde{F}^{\ominus 3})] \,  \mathcal{A}_{n}^{I} 
           - \frac{1}{2f} \frac{\partial f}{\partial\rho}
           \, 2\tilde{F}^{\ominus \boxplus}\, \mathcal{A}_{n}^{II}=0\,,\nonumber\\
&&[(\frac{\partial}{\partial\rho} - \frac{n}{\rho}) 
           - \frac{1}{2f} \frac{\partial f}{\partial\rho} 
           (1-2F_{56}-2\tilde{F}_{56}+ 2\tilde{F}^{\ominus 3})] \,  \mathcal{A}_{n}^{II} 
           -  \frac{1}{2f} \frac{\partial f}{\partial\rho}
           \, 2\tilde{F}^{\ominus \boxminus}\, \mathcal{A}_{n}^{I}=0\,,
\end{eqnarray}
and into equations for $\mathcal{B}_{n+1}^{I,II}$
\begin{eqnarray}
\label{massless12B}
&&[(\frac{\partial}{\partial\rho} + \frac{n+1}{\rho}) 
           - \frac{1}{2f} \frac{\partial f}{\partial\rho} 
                (1+2F_{56}-2\tilde{F}_{56} - 2\tilde{F}^{\ominus 3})] \, \mathcal{B}_{n+1}^{I} 
           - \frac{1}{2f} \frac{\partial f}{\partial\rho} 
           \, 2\tilde{F}^{\ominus \boxplus}\,  \mathcal{B}_{n+1}^{II}=0\,,\nonumber\\
&&[(\frac{\partial}{\partial\rho} + \frac{n+1}{\rho}) 
           - \frac{1}{2f} \frac{\partial f}{\partial\rho} 
              (1+2F_{56}-2\tilde{F}_{56}+2\tilde{F}^{\ominus 3})] \, \mathcal{B}_{n+1}^{II} 
           - \frac{1}{2f} \frac{\partial f}{\partial\rho} 
              \, 2\tilde{F}^{\ominus \boxminus}\, \mathcal{B}_{n+1}^{I}=0\,.
\end{eqnarray}
Both groups of equations can be solved in an equivalent way. 

Let us first find solutions for $\mathcal{A}_{n}^{I,II}$. In the case that $\tilde{F}^{\ominus \boxplus}$
$\tilde{F}^{\ominus \boxminus} = 0 =$ $\tilde{F}^{\ominus 3}$ the two functions decouple 
and one easily finds the solution~\cite{hn,dhn}:
$\mathcal{A}_{n}^{I,II}= {\cal N} \rho^{n}\, f^{\frac{1}{2}(1- 2F_{56} -2
\tilde{F}_{56})}$. 
In our case we correspondingly try with the following  ansatz
\begin{eqnarray}
\label{massless12Asol}
\mathcal{A}_{n}^{I }&=& 
\rho^{n}\, f^{\frac{1}{2}(1- 2F_{56} -2\tilde{F}_{56})} [a_1 f^{\alpha} + b_1 f^{\beta}]\,,\nonumber\\ 
\mathcal{A}_{n}^{II}&=& 
\rho^{n}\, f^{\frac{1}{2}(1- 2F_{56} -2\tilde{F}_{56})} [a_2 f^{\alpha} + b_2 f^{\beta}]\,,
\end{eqnarray}
When using this ansatz in Eq.~(\ref{massless12A}) we end up with the  relations
\begin{eqnarray}
\label{massless12Asol1}
f^{\alpha}\,(-a_{1}\tilde{F}^{\ominus 3} - \alpha a_{1} + a_{2 }\tilde{F}^{\ominus \boxplus})=&0&= 
f^{\beta} \,(-b_{1}\tilde{F}^{\ominus 3} - \beta b_{1} +  b_{2 }\tilde{F}^{\ominus \boxplus})\,,\nonumber\\
f^{\alpha}\,(a_{2}\tilde{F}^{\ominus 3} - \alpha a_{2} + a_{1 }\tilde{F}^{\ominus \boxminus})=&0&= 
f^{\beta}\, (b_{2}\tilde{F}^{\ominus 3} - \beta b_{2} +  b_{1 }\tilde{F}^{\ominus \boxminus})\,.
\end{eqnarray}
We correspondingly find
\begin{eqnarray}
\label{massless1234AB}
a_{2}  &=& a_{1}\, \frac{\alpha+ \tilde{F}^{\ominus 3}}{\tilde{F}^{\ominus \boxplus}} 
        =  a_{1}\, \frac{\tilde{F}^{\ominus \boxminus}}{\alpha - \tilde{F}^{\ominus 3}} \,,\nonumber\\
\alpha &=& \pm \sqrt{(\tilde{F}^{\ominus 3})^2+\tilde{F}^{\ominus \boxplus} \tilde{F}^{\ominus \boxminus}}\,,
\nonumber\\
b_{2}  &=& b_{1}\, \frac{ \beta+ \tilde{F}^{\ominus 3}}{\tilde{F}^{\ominus \boxplus}} 
        =  b_{1}\, \frac{\tilde{F}^{\ominus \boxminus}}{\beta  - \tilde{F}^{\ominus 3}} \,,\nonumber\\
\beta  &=& \pm \sqrt{(\tilde{F}^{\ominus 3})^2+\tilde{F}^{\ominus \boxplus} \tilde{F}^{\ominus \boxminus}}
= \alpha \,.
\end{eqnarray}

General massless solutions of Eq.~(\ref{massless12A}) are then
\begin{eqnarray}
\label{masslessgen}
\mathcal{A}_{n}^{I }&=& a\,
\rho^{n}\, f^{\frac{1}{2}(1- 2F_{56} -2\tilde{F}_{56})}\,f^{\alpha} \,,\nonumber\\ 
\mathcal{A}_{n}^{II}&=&  a \, \frac{\alpha+ \tilde{F}^{\ominus 3}}{\tilde{F}^{\ominus \boxplus}}\,
\rho^{n}\, f^{\frac{1}{2}(1- 2F_{56} -2\tilde{F}_{56})} \,f^{\alpha} \,,
\end{eqnarray}
with $\alpha$ equal to either  $-\sqrt{(\tilde{F}^{\ominus 3})^2+
\tilde{F}^{\ominus \boxplus} \tilde{F}^{\ominus \boxminus}}$ or to $+\sqrt{(\tilde{F}^{\ominus 3})^2+
\tilde{F}^{\ominus \boxplus} \tilde{F}^{\ominus \boxminus}}$. One can take also a superposition of both, 
with related  $\mathcal{A}_{n}^{I }$  and $\mathcal{A}_{n}^{II }$ as required by 
Eqs.~(\ref{massless1234AB}, \ref{masslessgen}). 

In a similar way we get massless solutions of Eq.~(\ref{massless12B})
\begin{eqnarray}
\label{massless12Bsol3}
\mathcal{B}_{n+1}^{I }&=& b\,
\rho^{-n-1}\, f^{\frac{1}{2}(1 + 2F_{56} -2\tilde{F}_{56})}\,f^{\alpha} \,,\nonumber\\ 
\mathcal{B}_{n+1}^{II}&=& b\, \frac{\alpha+ \tilde{F}^{\ominus 3}}{\tilde{F}^{\ominus \boxplus}}\,
\rho^{-n-1}\, f^{\frac{1}{2}(1 + 2F_{56} -2\tilde{F}_{56})} \,f^{\alpha} \,,
\end{eqnarray}
again with $\alpha = \mp \sqrt{(\tilde{F}^{\ominus 3})^2+
\tilde{F}^{\ominus \boxplus} \tilde{F}^{\ominus \boxminus}}$.

A function $\mathcal{G}$, defined on an infinite disc with the vielbein~\cite{dhn} 
$f= (1+ \frac{\rho^2}{(2 \rho_{0})^2})$, is normalizable provided
\begin{eqnarray}
\label{normalizable}
\int_{0}^{\infty} \, \rho d\rho \, f^{-2} \, |\mathcal{A}_{n}^{I,II }  |^2 < \infty\,, \nonumber\\
\int_{0}^{\infty} \, \rho d\rho \, f^{-2} \, |\mathcal{B}_{n+1}^{I,II }|^2 < \infty\,. 
\end{eqnarray}
Correspondingly are the solutions $\mathcal{A}^{I,II}_{n}$  normalizable under the condition
\begin{equation}
\label{norm-solA12}
-1 < n < 2 (F_{56}+ \tilde{F}_{56} \pm \sqrt{(\tilde{F}^{\ominus 3})^2+ \tilde{F}^{\ominus \boxplus} 
\tilde{F}^{\ominus \boxminus} }\,)\,.
\end{equation} 
The equivalent condition follows from normalizability requirement for  $\mathcal{B}^{I,II}_{n} $  
\begin{equation}
\label{norm-solB12}
 2 \,(F_{56}- \tilde{F}_{56} \pm \sqrt{(\tilde{F}^{\ominus 3})^2 + \tilde{F}^{\ominus \boxplus}
 \tilde{F}^{\ominus \boxminus}})< n <1 \,. 
\end{equation}
To find massive solutions one could follow the procedure of the ref.~\cite{dn}.


\begin{thebibliography}{99}
%
\bibitem{DNFAM} D. Lukman, N. S. Manko\v c Bor\v stnik, "Families of spinors in $d=(1+5)$ with 
                a zweibein and two kinds of spin connection fields on  an almost $S^2$", in 
                this Proceedings.
                               %
\bibitem{dhn} D. Lukman, N. S. Manko\v c Bor\v stnik, H. B. Nielsen, "Families of Spinors in $d=(1+5)$ 
		               Compactified on an Infinite Disc with the Zweibein Which Makes a Disc Curved 
		               on $S^2$ and a Possibility for Masslessness", 
                 p. 193-202,  arXiv:1012.0224.     
%
\bibitem{hn} D. Lukman, N. S. Manko\v c Bor\v stnik, H. B. Nielsen, "An effective two 
               dimensionality" cases bring a new hope to the Kaluza-Klein-like theories'', 
               http://arxiv.org/abs/1001.4679v5,
               {\it New J. Phys.} {\bf 13} (2011) 103027. arXiv:1001.4679v4.
\bibitem{dn} D. Lukman, N. S. Manko\v c Bor\v stnik, "Spinor states on a curved infinite disc 
                   with non-zero spin-connection fields", http://arxiv.org/abs/1205.1714, {\it J. of Phys.A: 
               Math. Theor. }{\bf 45} (2012) 465401 (19pp), doi:10.1088/1751-8113/45/46/465401.
%
\end{thebibliography}
\end{document}